\documentclass{eptcs}
\providecommand{\event}{QPL 2012}

\usepackage{graphicx}
\usepackage{mathtools}
\usepackage{amssymb}
\usepackage{amsmath}
\usepackage{amsthm}
\usepackage{subfigure}        % this is used...

\newtheorem{prop}{Proposition}%[section]
\newtheorem{cor}{Corollary}%[section]
\theoremstyle{definition}
\newtheorem{defin}{Definition}%[section]
\theoremstyle{remark}
\newtheorem*{rem}{Remark}
\newcommand{\ket}[1]{\mbox{$|#1\rangle$}}

\title{The expressive power of quantum walks in terms of language acceptance}
\author{Katie Barr \qquad\qquad Viv Kendon
\institute{School of Physics and Astronomy, E. C. Stoner Building, University of Leeds, Leeds LS2 9 JT, UK}
\email{v.kendon@leeds.ac.uk}
}
\def\titlerunning{Quantum walks and language acceptance}
\def\authorrunning{K. Barr \& V. Kendon}
\begin{document}
\maketitle

\begin{abstract}
Discrete time quantum walks are known to be universal for quantum computation. This has been proven by showing that they can simulate a universal quantum gate set. In this paper, we examine computation by quantum walks in terms of language acceptance, and present two ways in which discrete time quantum walks can accept some languages with certainty. These walks can take quantum as well as classical inputs, and we show that when the input is quantum, the walks can also be interpreted as performing the task of quantum state discrimination.
\end{abstract}

%%%%%%%%%%%%%%%%%%%%%%%%%%%%%%%%%%%%%%%%%%%%%%%%%%%%%%%%%%%
\section{Introduction}
\label{sec:Intro}

Quantum walks are the quantum generalisation of random walks on discrete structures \cite{Aharonov93,Aharonov01}. Both continuous and discrete time quantum walks are known to be Turing universal \cite{Lovett10,Childs09,Childs12}.  This has been proven by showing that in both cases an elementary universal gate set can be simulated by a quantum walk. This maps the quantum walk onto the quantum circuit model.  Both types of walks propagate quantum amplitude deterministically along ``wires'' punctuated by gates formed using an appropriate combination of graph structures.  The discrete time walk additionally uses a set of coin operations designed to produce the required propagation.  Despite much effort, quantum circuits have not yet been realised experimentally beyond a few qubits.  The best quantum computation devices currently in existence are based on liquid state nuclear magnetic resonance (NMR) \cite{NMR}.  Other models of computation fit NMR devices better, in particular, the Latvian quantum finite automaton (LQFA) \cite{Algebraic}, which has been specially designed for that purpose.  The computational capabilities of LQFAs are characterised in terms of language acceptance, prompting our investigation into language acceptance by other models of computation, specifically, discrete time quantum walks. Due to the construction in \cite{Lovett10}, we know that there must exist a mapping from a discrete time quantum walk onto any other Turing universal model of computation.  Hence, non-universal tasks such as language acceptance must be possible, and the interest lies in the details of how a quantum walk can be configured to do it. 

\textbf{This work:} we describe work in progress applying discrete time quantum walks to language acceptance. We show that there are multiple ways to do this, by exploring a range of small examples.  We provide two simple constructions of graphs on which quantum walks can recognise the languages $\mathcal{L}_{eq}$ and $\mathcal{L}_{ab}$. Furthermore, we show that exploiting explicitly quantum aspects of the walks allows for gains in efficiency. We introduce the concept of quantum inputs, and describe a preliminary investigation into the effect these have on word acceptance.  Our work made use of numerical simulations; the technical details of how these were done have been published in \cite{barr13b}.

The paper is arranged as follows: in 
%section \ref{sec:back} we give some further background and context to motivate our investigations.  In 
section \ref{sec:defs} we provide definitions covering the elements we use from formal languages and finite state automata; the Jaro distance; and discrete time quantum walks. In section \ref{sec:qw2fl}, we give examples of languages which can be accepted by quantum walks using two different graph formats and discuss their efficiency.  In section \ref{sec:qinputs}, we note that the constructions permit quantum inputs, and briefly indicate how these might be used to apply the quantum walks to quantum state discrimination \cite{Chefles00, QWPOVM}.  We concludes with a summary and outline of future work in section \ref{sec:sum}.

%%%%%%%%%%%%%%%%%%%%%%%%%%%%%%%%%%%%%%%%%%%%%%%%%%%%%%%%%%%
\section{Definitions and notation}
\label{sec:defs}

Here we provide definitions used in the paper, thus setting up our notation.

\subsection{Languages}
\label{sec:FLdef}

We consider a binary alphabet $\Sigma$ with two symbols $\{a,b\}$, from which words are formed as strings of symbols. 
\begin{defin}
The language $\mathcal{L}_{ab}$ is the set of words $\{(ab)^n| n \in \mathbb{N}\}$.
\end{defin}
\begin{rem}
This is an example of a regular language.  Each word in $\mathcal{L}_{ab}$ is a different length.
\end{rem}

\begin{defin}
The language $\mathcal{L}_{eq}$ is the set of words $\{a^mb^m| m\in \mathbb{N}\}$.
\end{defin}
\begin{rem}
This is an example of a context-free language.  Recognising this language requires a basic form of memory, to count how many times the first symbol occurs and check the second symbols then occurs the same number of times.  However, like $\mathcal{L}_{ab}$, each word is a different length.
\end{rem}

%%%%%%%%%%%%%%%%%%%%%%%%%%%%%%%%%%%%%%%%%%%%%%%%%%%%%%%%%%%
\subsection{Jaro distance}
\label{sec:Jaro}

The Jaro distance is a metric which indicates how similar two strings (words) are. 
\begin{defin}
For strings $w_1$ and $w_2$, the \emph{match distance} $m_d$ is given by
\begin{equation}
m_d = \left\lfloor\frac{\text{max}(|w_1|,|w_2|)}{2}\right\rfloor -1
\end{equation}
\end{defin}

Two charcters in strings $w_1$ and $w_2$ are said to \emph{match} if
(1) they are the same symbol and (2) if their positions within their
respective strings lie within the match distance of each other.  Let
$s$ denote the number of matching characters.  Let $w'_1$ and $w'_2$
be the substrings derived from $w_1$ and $w_2$ respectively by erasing
all the non-matching characters.  Then $t$, the number of
\emph{transpositions}, is the number of positions at which $w'_1$ and
$w'_2$ differ.   Now we define the Jaro distance as follows.

\begin{defin}
For strings $w_1$ and $w_2$, the \emph{Jaro distance} $d_j$ between $w_1$ and $w_2$
is given by
\begin{equation}
d_j = \left\{ \begin{array}{cc} 0  & \text{if } s = 0 \\
\frac{1}{3} ( \frac{s}{|w_1|} + \frac{s}{|w_2|} + \frac{s-t}{s}   ) & \text{otherwise} \end{array} \right.
\end{equation}
where $s$ is the number of matching characters, i.e., characters which occur in both strings, in the same order, within the match distance $m_d$. The value of $t$ is obtained by dividing the number of matching characters which differ by sequence order by 2. 
\end{defin}
\begin{rem}
The three parts to the expression for the Jaro distance calculate the
ratios of the number of matching characters to the lengths of $w_1$
and $w_2$ and then the fraction of non-transpositions among the matching characters.  The Jaro distance was selected as it always has values between 0 and 1, with 1 indicating that two words are equal, hence it was easy to compare to the probability of acceptance.  Other metrics with similar properties for comparing strings could have been used; the Jaro distance was available as a Python module, which made it a convenient choice for the numerical part of the work.
\end{rem}

%%%%%%%%%%%%%%%%%%%%%%%%%%%%%%%%%%%%%%%%%%%%%%%%%%%%%%%%%%%
\subsection{Discrete time quantum walks}
\label{sec:discqw}

\begin{defin}
An undirected graph $G = \{E, V \}$ consists of a set of vertices $v\in V$ and edges $e\in E$ joining pairs of vertices.  The size of the graph $N$ is the number of vertices $N=|V|$.  The set of edges that meet at vertex $v$ is denoted $E_v$.  The degree $d_v$ of a vertex $v$ is the number of edges meeting at that vertex, $d_v = |E_v|$.  The maximum degree $d_{\mathrm{max}}$ is $\max_v \{d_v|v\in V\}$.  At each vertex $v$ we label the edges from $\{0\dots d_v-1\}$ in an arbitrary but fixed order.
\end{defin}
\begin{rem}
Each edge thus has two labels, one at each end, which will usually be different.  This is not the only way to set up a discrete time quantum walk on an arbitrary graph, but it is the most convenient way for numerical calculations \cite{Kendon06}.
\end{rem}
\begin{defin}
The Hilbert space of a discrete-time quantum walk on the graph $G$ is $\mathcal{H}_V\times\mathcal{H}_{d_{\mathrm{max}}}$, spanned by basis states $\ket{v,c}$, $v\in V$ and $c\in E_v$.
A state of the quantum walk is written as superposition of basis states,
$\ket{\psi} = \sum_{v,c}\beta_{v,c}\ket{v,c}$ with amplitude $\beta_{v,c}\in \mathbb{C}$ and with normalisation $\sum_{v,c}|\beta_{v,c}|^2 = 1$.
\end{defin}
\begin{defin}
A single-vertex coin operator $C_v$ is a $d_v$ by $d_v$ unitary operator.  A coin operator for the whole graph is the direct sum of single-vertex coin operators, $C=\sum_v C_v$.  
\end{defin}
\begin{rem}
The dimensions of this square matrix $C$ are given by $\sum_v d_v$.
\end{rem}
\begin{defin}
The shift operator $S$ is a unitary operator that translates the amplitude $\beta_{v,c}$ to the vertex $v'$ that is connected to $v$ by edge $c$, i.e., $S\ket{v,c} = \ket{v',c'}$, where $c'$ is the label of this edge at $v'$.
\end{defin}
\begin{defin}
A \emph{discrete time quantum walk} (DQW) on the graph $G$
starting in initial state $\ket{\psi_0}$ evolves according to a combined coin and shift unitary operator $U = SC$.  The state $|\psi(T)\rangle$ of the DQW after $T$ steps is $|\psi(T)\rangle = U^T\ket{\psi_0}$.  The probability $P(v,T)$ of finding the walker at a chosen vertex $v$ at time $T$ is
\begin{equation}
P(v,T) = \sum_{c\in v} |\beta_{c,v}(T)|^2 \label{eqn:vprob}
\end{equation}
i.e., the sum of the square moduli of the amplitude for each edge at vertex $v$.
\end{defin}

\begin{figure}
\centering 
\begin{minipage}{0.6\textwidth}
\includegraphics[scale=0.3]{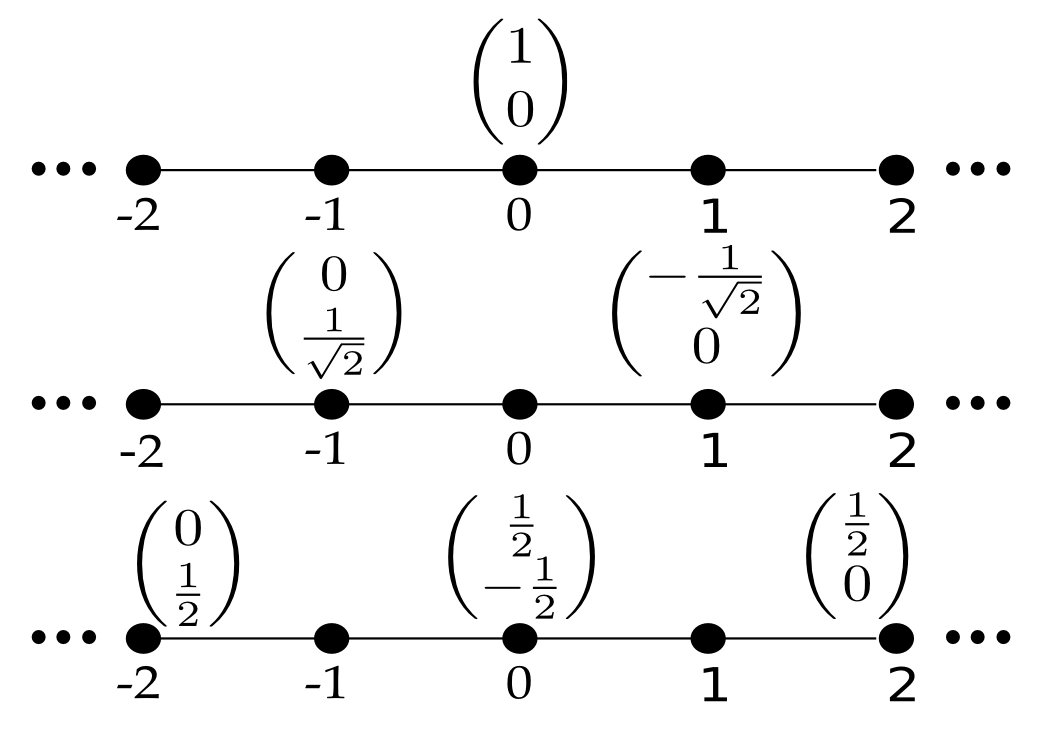}
\end{minipage}
\begin{minipage}{0.1\textwidth}
$T=0$\\[4em]
$T=1$\\[4em]
$T=2$
\end{minipage}
\caption{Two steps of a discrete time quantum walk on the line, starting at the origin, evolved with a Hadamard coin operator. The numbers in parentheses are the amplitudes for each basis state, one for each edge at each vertex.}\label{fig:line}
\end{figure}
\begin{rem}
The discrete time quantum walk is the quantum analogue of the classical random walk, in the sense that if you measure the quantum walk after each step, the quantum behaviour is 
converted to a classical random walk.  As an example, the quantum walk on
the line is shown in figure \ref{fig:line}.   The vertices at the integer points on the line have degree two, so we use a coin operator given by the Hadamard operator (the only non-trivial type of degree two \cite{Ambainis01}),
\begin{equation}
H = \frac{1}{\sqrt{2}} \begin{pmatrix} 1 & 1 \\ 1 & -1 \end{pmatrix} \label{eqn:hadamard}
\end{equation} 
\end{rem}

%%%%%%%%%%%%%%%%%%%%%%%%%%%%%%%%%%%%%%%%%%%%%%%%%%%%%%%%%%%
\section{Applying quantum walks to language recognition}
\label{sec:qw2fl}

To set up a discrete time quantum walk on a graph to differentiate between words, and hence determine whether they are in a given formal language, we need to
specify suitable encodings of input and output that correspond with the graph structure and
the quantum state of a quantum walk on that graph.  The problem is thus under-specified:
there are many possible graphs we could use and many ways to encode a sequence of symbols into a quantum state of a quantum walk on a graph.  First of all, we will restrict ourselves to single quantum walkers.  Although efficient universal quantum computation with quantum walks requires multiple walkers \cite{Childs12}, our goal here is to find simple examples of non-universal computation, for which we can expect a single walker will be sufficient \cite{Lovett10,Childs09}, though not necessarily efficient.  

We therefore first chose a simple encoding of a binary alphabet that will work for a single quantum walker.  A \emph{dual rail encoding} for the input word assigns each symbol to a pair of vertices.  The initial state will have amplitude on the first vertex of the pair for the first symbol in the alphabet, and vice versa for the second symbol, see figure \ref{fig:spasetup}.  A very similar dual rail encoding can instead assign each symbol to one of a pair of edges at a vertex, see section \ref{sec:sequ}.  Since quantum computations should start with an initial state that is easy to prepare, or else we could be hiding significant extra computation in the state preparation, we must check that such an input state is experimentally feasible.  While it would require a multiport interferometer to split the quantum state over the different vertices, modulated to enable differentiation between $a$ and $b$ inputs, a dual rail encoding would be straightforward to configure for any given sequence of symbols.

Next, a graph structure and coin operations which transform the input into an appropriate output must be chosen. There are a variety of ways that this can be done. For example, if we use a single walker starting in a superposition that represents the input distributed along different vertices, the entire word can be operated on simultaneously.  Alternatively, the superposition representing the input symbols can be fed into the graph structure one symbol after another.   We will examine each of these in turn in the next two subsections.

The simplest way to deal with the output appears to be to designate an ``accepting vertex'', which all accepting amplitude should be arrive at simultaneously. Measuring the position of the quantum walker after the appropriate number of time steps will then provide the output of the computation.  If the position is found to be the accepting vertex, the quantum walk has accepted the input word.  The outcome of a quantum measurement is in general not deterministic: we shall therefore use the following definition of acceptance for our quantum walk language recognition. 
\begin{defin}\label{def:acbe}
The language $\mathcal{L}$ recognised with cut-point $\lambda \in [0, 1)$ by a DQW is $\mathcal{L}=\{w|w \in\Sigma^* \, P(w)> \lambda \}$ where $P(w)$ is the probability measuring the DQW at the accepting node for input word $w$. The acceptance by a DQW of $\mathcal{L}$ is with \emph{bounded error} if there is some $\epsilon> 0$ such that for all $w \in\mathcal{L}$, the DQW accepts $w$ with probability greater than $\lambda + \epsilon$ and any words $w \notin \mathcal{L}$ are accepted with probability less than $\lambda - \epsilon$ where $\epsilon$ is the error margin. If there is no such $\epsilon$ then the DQW accepts $\mathcal{L}$ with \emph{unbounded error}.
\end{defin}
\begin{rem}
This is based on the standard definition of acceptance for finite state automata and languages.  
\end{rem}

This can easily be generalised to a set of accepting vertices; finding the walker on any one of them indicating acceptance.  The problem then is to find graph structures and sets of coin operations which transport a large proportion of the amplitude to the accepting vertex or vertices when the input is a word from the language the walk is designed to accept.  If the remaining amplitude can be redirected to another vertex, the rejecting vertex, then the accepting and rejecting conditions can be inverted to allow the same walk to accept both the language and its complement -- the set of words not in that language.   By transferring the input superposition to an output vertex, the language acceptance problem can thus be seen as a variation of the the quantum state transfer problem, see, for example \cite{PSTrev, Bose03, Christandl05}.

Some simple cases are immediately evident.
The empty set and empty string are accepted trivially, as we can distinguish between no walk occurring and all amplitude being rejected. Singleton symbols can be accepted by \emph{paths} of length three with the amplitude initially in the appropriate coin state of the central vertex and swap operators directing all the amplitude to one end vertex or the other, depending on which of the two singleton symbols is encoded.  These are too trivial to make use of any quantum properties of the quantum walk, we simply note them for completeness.

%%%%%%%%%%%%%%%%%%%%%%%%%%%%%%%%%%%%%%%%%%%%%%%%%%%%%%%%%%%
\subsection{Spatially distributed input}
\label{sec:spatial}

\begin{figure}[t]
\centering \includegraphics[scale = 0.25]{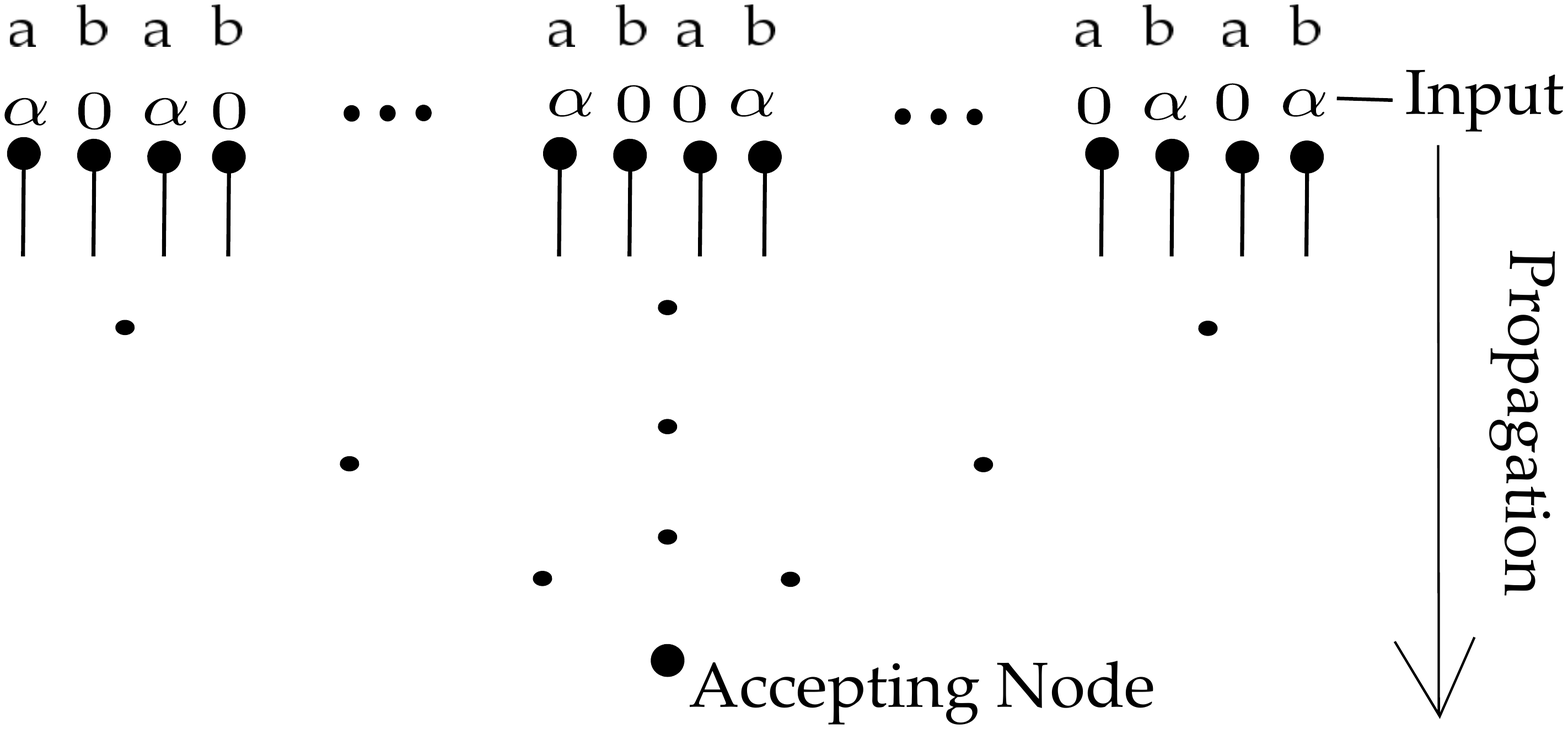} 
\caption{Schema for the graph structure required by a quantum walk which accepts languages using a spatially distributed input.  The initial state of the quantum walker is equally distributed over the vertices labeled $\alpha$, with vertices labeled $0$ being unoccupied.  The corresponding symbols $a$ and $b$ are shown above the vertices, in this case the input is $aa...ab...bb$.} \label{fig:spasetup}
\end{figure}

We will first consider the case where the input is prepared in a dual rail encoding across a row of vertices of the graph structure, the general form of which is shown schematically in figure \ref{fig:spasetup}. There are two vertices for each input symbol with alternate vertices representing $a$ or $b$. The length $n$ of the input must be known. For each input symbol $s_j$, $0<j\le n$, the vertex $2j-1$ is populated with amplitude $1/\sqrt{n}$ if $s_j=a$ else $2j$ contains amplitude if $s_j=b$. The structure of the graph and coin for a walk testing whether a word is in a given language depends on $n$. The walk is then run for a predetermined number of steps. When the initial state encodes a word in the language accepted by the walk, the amplitude should be directed to the designated accepting vertex. If a word not in the language accepted is encoded in the initial state then less of the amplitude should be directed towards the accepting vertex, so the acceptance will be with bounded error (Definition \ref{def:acbe}). The modulus squared of the final amplitude at the accepting vertex, equation (\ref{eqn:vprob}) yields the probability of acceptance.

\begin{figure}[t]
\centering \subfigure (a) \includegraphics[scale = 0.185]{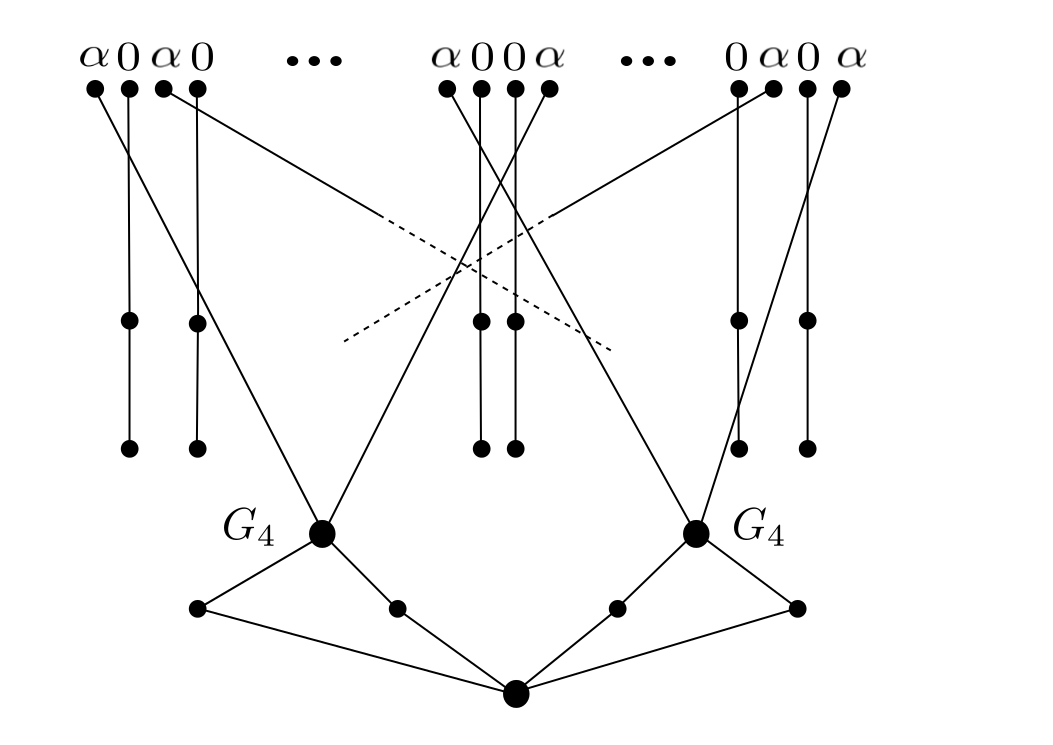} \subfigure (b) \includegraphics[scale = 0.23]{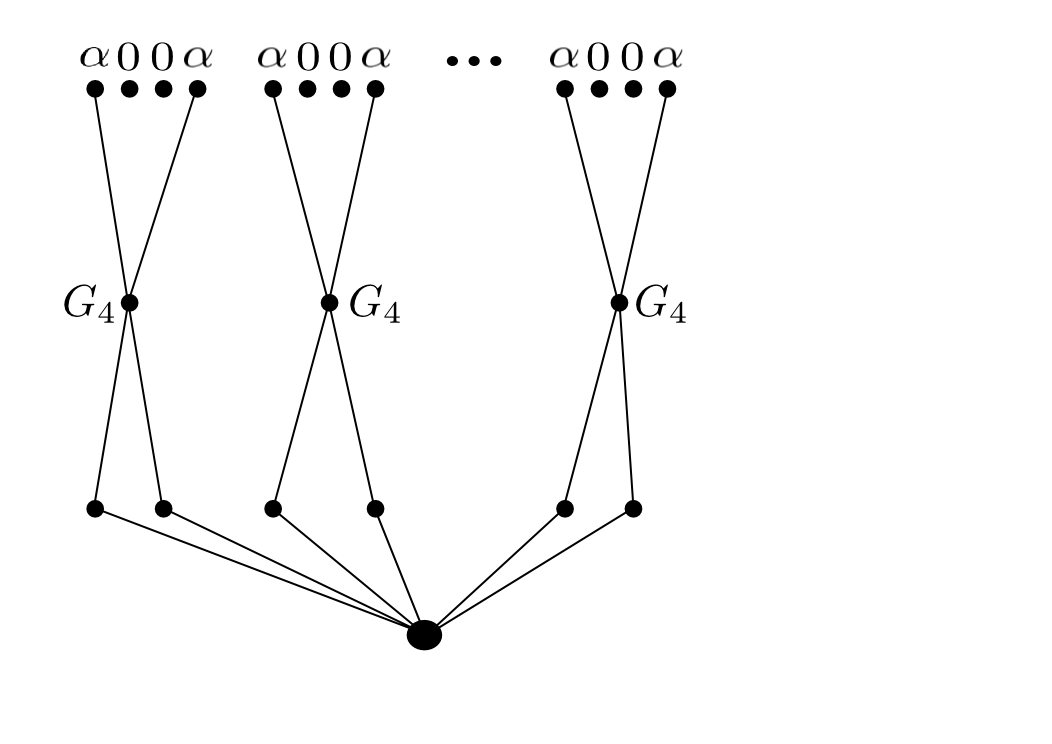}
\caption{Graphs on which quantum walks accept a) $\mathcal{L}_{eq}$ and b) $\mathcal{L}_{ab}$ with each input symbol being operated on simultaneously. Edges join only at vertices indicated by black circles.  Grover coins of the appropriate dimensions are used at each vertex, and only two of the $G_4$ vertices are shown -- there are $m$ in total for words of length $2m$, with the $j$th $G_4$ connected to the inputs representing the $j$th $a$ and $j$th $b$.  These graphs are used in the proof of Proposition \ref{proof} and Corollary \ref{cor}.}
\label{fig:ambmspa2}
\end{figure}

This design is most easily used to swiftly accept languages which contain at most one word of each length, so the quantum walk graph structure tests for that specific word. We use the examples $\mathcal{L}_{eq} = \{ a^m b^m | m \in \mathbb{N}\}$ and $ \mathcal{L}_{ab}= \{ (ab)^m | m \in \mathbb{N} \}$ and the graphs accepting them can be seen in figure \ref{fig:ambmspa2}. 

The constructions are based on a $d$ dimensional Grover operator:
\begin{equation}
\label{eqn:grover}
G_{d}= \begin{pmatrix}
 \frac{2-d}{d} & \frac{2}{d} & \cdots & \frac{2}{d} \\
& & &\\
  \frac{2}{d} & \frac{2-d}{d} & \cdots & \frac{2}{d} \\
& & & \\
  \vdots  & \vdots  & \ddots & \vdots  \\
& & & \\
   \frac{2}{d}  & \frac{2}{d} & \cdots & \frac{2-d}{d}
 \end{pmatrix}
\end{equation}
where $d$ is the degree of the vertex.
With $d$ even, if $d/2$ of the edges contain amplitude equally distributed between them, this coin operator transfers all the amplitude to the other $d/2$ edges. The shift operator then moves the amplitude to the connected vertices.   The deterministic evolution that the Grover operator can produce was used to design the ``wires'' which transmitted the amplitude between gates in \cite{Lovett10}.

\begin{prop}
\label{proof}
The language $\mathcal{L}_{eq}$ is accepted with certainty by the graph and choice of operators shown in figure \ref{fig:ambmspa2} (a). Words not in the language are accepted with bounded error. 
\end{prop}

\begin{proof}
\textit{Words in $\mathcal{L}_{eq}$:} The result follows by induction on $m$. For the base step, simple calculation and equation \ref{eqn:grover} shows that $ab$ will be accepted with certainty. For the induction step, suppose that a word $a^m b^m$ is accepted with certainty, and consider the word $a^{m+1} b^{m+1}$. Due to the construction of the graph, every pair $a_i b_{i+m}$ where $i \leq m$ is treated independently, so by the induction hypothesis all amplitude from the first $m$ $a$'s and $b$'s is transmitted to the accepting vertex. The calculation to show that all amplitude from the subsequent pair of $a$'s and $b$'s is identical to the base step.\\
\textit{Words not in $\mathcal{L}_{eq}$:} As there is no path between the input vertices representing symbols which do not occur in words from $\mathcal{L}_{eq}$, their amplitudes cannot contribute to the accepting probability. To prove the acceptance is with bounded error, consider a word which differs from a word in  $\mathcal{L}_{eq}$ by one symbol, for example $a^m b^{m-1} a$. This will be accepted with the maximum possible probability for a word not in $\mathcal{L}_{eq}$. The amplitude from the final $a$ cannot be transmitted to the accepting vertex, so the word cannot be accepted with probability $> 1 - \frac{1}{2m}$. Additionally, after step one of the walk the $m$'th $a$ goes to the Grover operator, and now some of this amplitude will be transmitted back to the input vertex:

\begin{equation}
G_4 \begin{pmatrix} \frac{1}{\sqrt{2}n} \\ 0 \\ 0 \\ 0 \end{pmatrix} = \begin{pmatrix} - \frac{1}{2} & \frac{1}{2} & \frac{1}{2} & \frac{1}{2} \\ \frac{1}{2} & - \frac{1}{2} & \frac{1}{2} & \frac{1}{2} \\ \frac{1}{2} & \frac{1}{2} & - \frac{1}{2} & \frac{1}{2} \\ \frac{1}{2} & \frac{1}{2} & \frac{1}{2} & -\frac{1}{2}\end{pmatrix} \begin{pmatrix} \frac{1}{\sqrt{2m}} \\ 0 \\ 0 \\ 0 \end{pmatrix} = \begin{pmatrix} -\frac{1}{2\sqrt{2m}} \\ \, \frac{1}{2\sqrt{2m}} \\ \, \frac{1}{2\sqrt{2m}} \\ \, \frac{1}{2\sqrt{2m}} \end{pmatrix}
\end{equation} 

Hence the total probability of accepting $a^m b^{m-1} a$ is $1-  \frac{1}{8m} - \frac{1}{4m}$.
\end{proof}

\begin{center}
\begin{figure}[t]
\centering \includegraphics[scale = 0.4]{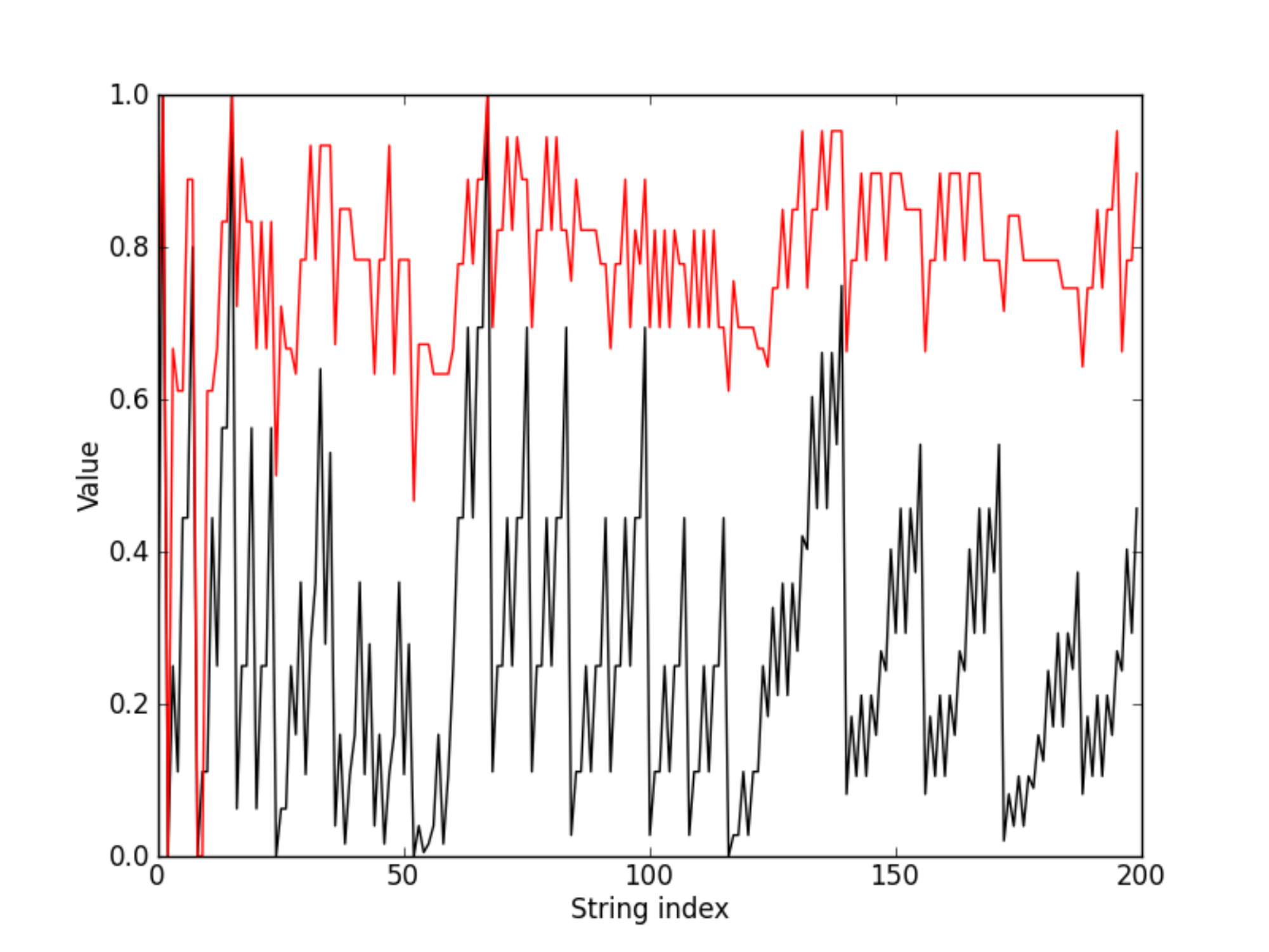}
\caption{The probability of acceptance for the walk detecting words from $\mathcal{L}_{eq}$ for the first 200 words (black). The Jaro distance between the input word and an appropriately sized word from $\mathcal{L}_{eq}$ is indicated in red. Both curves go to unity at the positions representing the words $ab$, $aabb$ and $aaabbb$.} 
\label{fig:ambmspa}
\end{figure}
\end{center}

\begin{cor}
\label{cor}
The language $\mathcal{L}_{ab}$ is accepted with certainty by the graph and choice of operators shown in figure \ref{fig:ambmspa} (b).
\end{cor}

In both of these walks, the input is processed in three full steps of the walk, regardless of the length of the input word. However the $O(1)$ time complexity is at the expense of $O(n)$ spatial complexity. For inputs of length $n$, the walks accepting $\mathcal{L}_{eq}$ and $\mathcal{L}_{ab}$ require  $4n + 3$ and $4n + 1$ vertices respectively. 
The Grover operator can also be exploited to generate quantum walks accepting with certainty other languages such as $\mathcal{L}_{twin} = \{ ww | w \in \{a, b\}^* \} $ and $\mathcal{L}_{rev} = \{ ww^r | w \in \{a, b\}^* \} $ where $w^r$ denotes the symbols of $w$ in reverse order.
The graph from figure \ref{fig:ambmspa2} a) can be extended to accept the archetypal context-sensitive language $\mathcal{L} = \{ a^m b^m c^m | m \in \mathbb{N} \}$ and b) to accept $\mathcal{L} = \{ (abc)^m | m \in \mathbb{N} \} = \{ abc \}^*$. In this case we must extend the model to deal with more than two input symbols which will involve a corresponding increase in the number of vertices required.  

\textbf{Numerical study:}
the properties of the walks over the graphs depicted in figure \ref{fig:ambmspa2} were investigated by simulating them using the Python programming language for all possible inputs of a given length, up to $n=16$. As well as the acceptance probability, the Jaro distance between the input and the closest word to a word in the language under consideration for that length was also calculated.  In the case of even length inputs, the comparison is made with the word from the language of that length, for odd inputs with length $n$ the word was compared to the word in the language of length $n-1$.  
For both languages the results were very similar so we limit the discussion to $\mathcal{L}_{eq}$ here. 
The results for the first 200 words are illustrated in figure \ref{fig:ambmspa}. The points at which both curves peak at the value one are at the position of the words $ab$, $aabb$, and $aaabbb$. 

By numerically comparing the probability of accepting an arbitrary word to a measure of how similar that word is to one of the required form, the algorithm's correctness is shown, and more detail concerning the behaviour of the walk on words not in the language  can be ascertained. The disparity between the Jaro distance and the probability of acceptance for words not in $\mathcal{L}_{eq}$ illustrates how effective the quantum walk algorithm is. A low probability of acceptance for any word not in the language, regardless of how close that word is to a word in $\mathcal{L}_{eq}$ indicates the language acceptance is robust. Hence this probability cannot be used as a good measure of how close the input word is to one in that language in cases where it is not equal to one. 

Using a spatially distributed input allows for long words to be accepted with the same number of operations as short words. However, the number of vertices required in the graph structure grows, albeit linearly, with the length of the word. The number of vertices required to accept a given language can be held constant regardless of input length if each input symbol is fed into the structure in turn, so we now turn to this case.

%%%%%%%%%%%%%%%%%%%%%%%%%%%%%%%%%%%%%%%%%%%%%%%%%%%%%%%%%%%
\subsection{Sequentially distributed input}
\label{sec:sequ}

\begin{figure}[h] \subfigure a) \includegraphics[scale = 0.29]{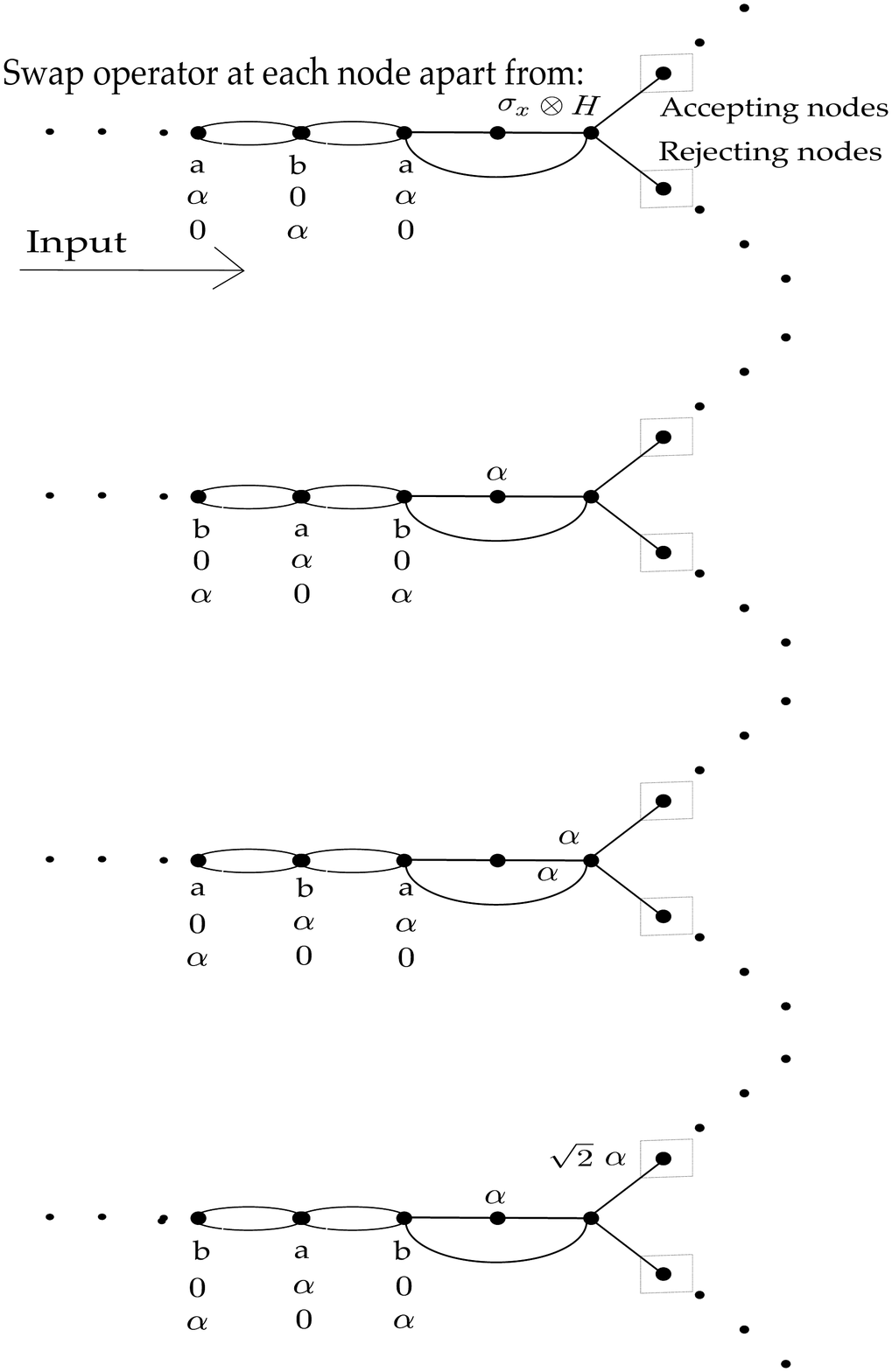} \subfigure b) \includegraphics[scale=0.29]{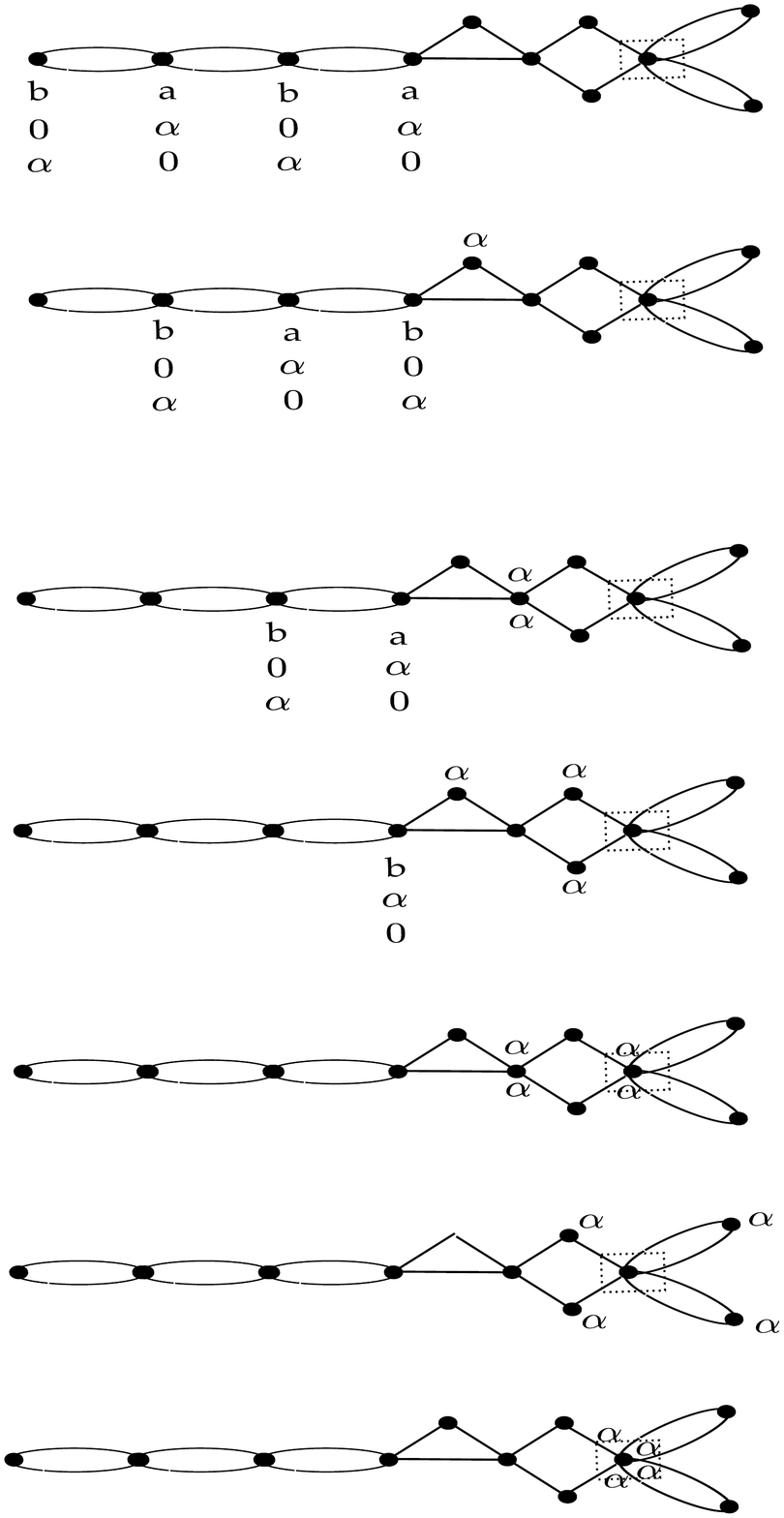}
\caption{a) Graph structure and coins accepting language $\mathcal{L}_{ab}$ b) Graph specifically accepting the word of length 4 from that language} 
\label{fig:abmstr}
\end{figure}

We can make use of the ``wires'' from \cite{Lovett10} with a slightly different encoding to design a sequential input.  An input of length $n$ can be treated sequentially if we start with it distributed along a chain of length $n$ with two links between each vertex, as shown in the leftmost portions of the graphs in figure \ref{fig:abmstr}. The two symbols are represented thus:
\begin{equation} 
a \Rightarrow \begin{pmatrix} \alpha \\ 0  \\ 0 \\ 0 \end{pmatrix} \,\,\,\,\,\,\,\,\,\,\,\,\,\,\,\,\,\,\,  b \Rightarrow \begin{pmatrix} 0 \\ \alpha \\ 0 \\ 0 \end{pmatrix}  ~\label{eqn:abs}
\end{equation}
where the entries in the vector contain the amplitude for each of the four edges at the vertex, before the coin operator is applied.
The coin on this part of the graph is $ \sigma_x \otimes \mathbb{I}_2$, where $\sigma_x$ is
the Pauli-X operator, 
\begin{equation}
\sigma_x =  \begin{pmatrix} 0 & 1 \\ 1 & 0 \end{pmatrix} \label{eqn:sigmax}.
\end{equation} 
This four-dimensional operator $ \sigma_x \otimes \mathbb{I}_2$ simply swaps amplitude to the ``leaving'' edges of the current vertices, the lower pair of entries.   The swap operator then transfers all the amplitude to the ``arriving'' edges of the next vertex. This choice of coin operator feeds the input amplitude into a graph which has either  ``accepting paths'' and ``rejecting paths'', or ``accepting vertices'' and ``rejecting vertices''. The sum of the square moduli of the amplitudes on the accepting vertex/path after a number of steps determined by $n$ gives the probability of accepting the input word. The shape of the graph and the coins at each vertex determine which words will be accepted. 

\textbf{Results:}
in some cases, such as walks accepting specific words of known length, the only coins required are trivial swap operators. More complex languages require less trivial coin operators, such as in figure \ref{fig:abmstr} (a), the graph accepting the language $\mathcal{L}_{ab}$. This graph uses the Hadamard operator, defined in equation (\ref{eqn:hadamard}), to determine whether each pair of symbols is of the form $ab$ or not. The swap operator then moves the amplitude from both symbols into the accepting path if they are of that form. Words in the language are accepted with certainty and those not in the language are accepted with probability $1/2$. 
Figure \ref{fig:abmstr} illustrates two different graphs which both accept the same specific word. Either a specific graph or the graph accepting $ \mathcal{L}_{ab} $ can be used to accept $abab$, as shown in figure \ref{fig:abmstr} (b) and (a) respectively. We tested numerically  whether these different graphs accepting the same word give rise to different probability distributions.  In this case, the probability of acceptance of words not equal to $abab$ did not depend on the graph, however, this property is not generally expected to hold. 
We can also see that exploiting quantum properties, using the Hadamard operator as in figure \ref{fig:abmstr} (a) to control interference between different parts of the amplitude, rather than permuting amplitude so that it all arrives at the right place eventually as in figure \ref{fig:abmstr} (b), gains us efficiency. The simple swapping version of the graph accepting $abab$ has $8$ vertices (discounting the input vertices) and takes $6$ steps to accept the word. The more general graph not only accepts more words, but accepts $abab$ in $5$ steps. To accept longer words from $\mathcal{L}_{ab}$ using a permutation scheme rather than the graph using the Hadamard operator, more steps are required and the size of the graph increases accordingly. The correctness of these quantum walk algorithms can be proven easily by induction on $m$, where $m$ is the number of times the string $ab$ is repeated. 

This scheme can be modified to accept $\mathcal{L}_{eq}$. This is because in the quantum walk shown in figure \ref{fig:abmstr} a), if a pair $ab$ was contained in the input, the amplitude from the $a$ and $b$ symbols arrives at the vertex where the $\sigma_x \otimes H$ at the same time, and interference can then occur. If the amplitude from the $a$ symbol is instead directed to a path of $m$ vertices prior to the vertex where $\sigma_x \otimes H$ is applied, then interference cannot occur until $m$ steps of the walk have taken place. If the $m+1$'th symbol of the input is then a $b$, then constructive interference will occur. Therefore if there are $m$ $a$'s followed by $m$ $b$'s  in the input, it will be accepted with certainty by the modified graph.  The graph now requires $O(n)$ vertices besides the input vertices, reflecting the need for a simple memory to recognise a context free language.

\begin{figure}[t]
\centering \includegraphics[scale = 0.4]{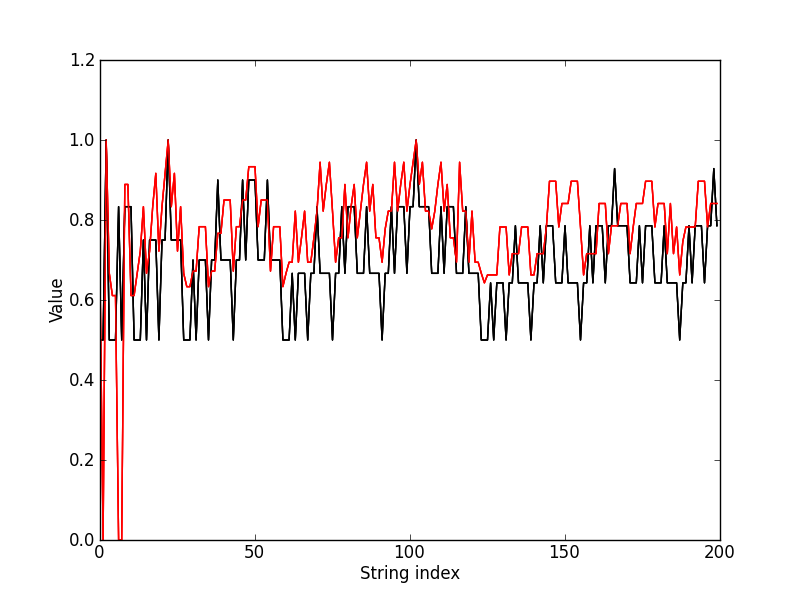}
\caption{Probability of acceptance using the graph from figure \ref{fig:abmstr} (a) accepting $\mathcal{L}_{ab}$ for the first 200 words (black). The Jaro distance between the input word and an appropriately sized word from the language is indicated in red.} 
\label{fig:abmseq}
\end{figure}

\textbf{Numerical study:}
as for the spatially distributed input case, the probability of accepting each word was calculated for all possible inputs up to length 16. This was plotted alongside the Jaro distance from the input word to a word of an appropriate length in the language as shown in figure \ref{fig:abmseq}. As in figure \ref{fig:ambmspa}, points at which both values go to unity indicate the positions of the word $ab$, $abab$, $ababab$. 
Although acceptance is still with bounded error,
comparing with figure \ref{fig:ambmspa} we see that the rejection of words not in the language is not so robust, all words are accepted with probability at least a half.

A simple way of finding a graph which accepts the string $abb$ concatenated onto a single symbol given either $aabb$ or $babb$ is to add further self loops (with a coin state for each end) to the accepting node.  A $b$ symbol requires two self loops for each portion of the existing amplitude to go around, and an $a$ symbol requires one self loop for each portion of the existing amplitude to go around. The rejecting paths leaving the accepting node will have to be increased accordingly. Whilst this will accept the new word with probability 1, if the process is repeated it will increase the probability of accepting other strings dramatically, though it will never be equal to unity.

%%%%%%%%%%%%%%%%%%%%%%%%%%%%%%%%%%%%%%%%%%%%%%%%%%%%%%%%%%%
\section{Quantum Inputs}
\label{sec:qinputs}

The inputs used so far have all been classical, represented by a quantum superposition state. However, the way these walks have been set up allows us to do more than this. Each symbol in the input word can be in a superposition of $a$ or $b$, for example, $ x |a\rangle + y |b\rangle $ such that $ |x|^2 + |y|^2 = \alpha^2 $.  Recall that $\alpha = 1/\sqrt{n}$ and $n$ is the length of the input. Superpositions of words, such as $abab$ and $bbbb$ can then be created by using the appropriate superposition for each symbol in the word. Where symbols match, the amplitude is allocated to that symbol as for the classical encoding.  Where symbols do not match, the amplitude is distributed between the $a$ and $b$ states accordingly. 
\begin{defin} 
A $\eta$-quantum input $|s^{1,2}\rangle$ is a superposition of the symbols in words $w_1$ and $w_2$ with amplitude ratios $\eta$ and $\sqrt{1-|\eta|^2}$. Where symbols of $w^1$ and $w^2$ match, so $w^1_i = w^2_i$, $|s^{1,2}_i \rangle = | w^1_i \rangle$. Where symbols differ $| s^{1,2}_i \rangle = \alpha(\eta | w^1_i \rangle +  \sqrt{1-|\eta|^2} | w^2_i \rangle$).
\end{defin} 
\begin{rem}
This is \emph{not} the same as a superposition of the whole words, for example, with $\eta=1/\sqrt{2}$,
\begin{equation}
(|aba\rangle + |bbb\rangle)/\sqrt{2} \ne (|a\rangle + |b\rangle)|b\rangle(|a\rangle + |b\rangle)/2 = (|aba\rangle + |abb\rangle + |bba\rangle + |bbb\rangle)/2.
\end{equation}
We chose to begin with symbol-by-symbol superpositions because they are simple to prepare.  Preparing superpositions of whole words would require more resources, so while they are undoubtedly also interesting, the preparation resources would need to be accounted for in the overall assessment of the algorithm.
\end{rem}

\begin{figure}[t]
\centering \includegraphics[scale = 0.4]{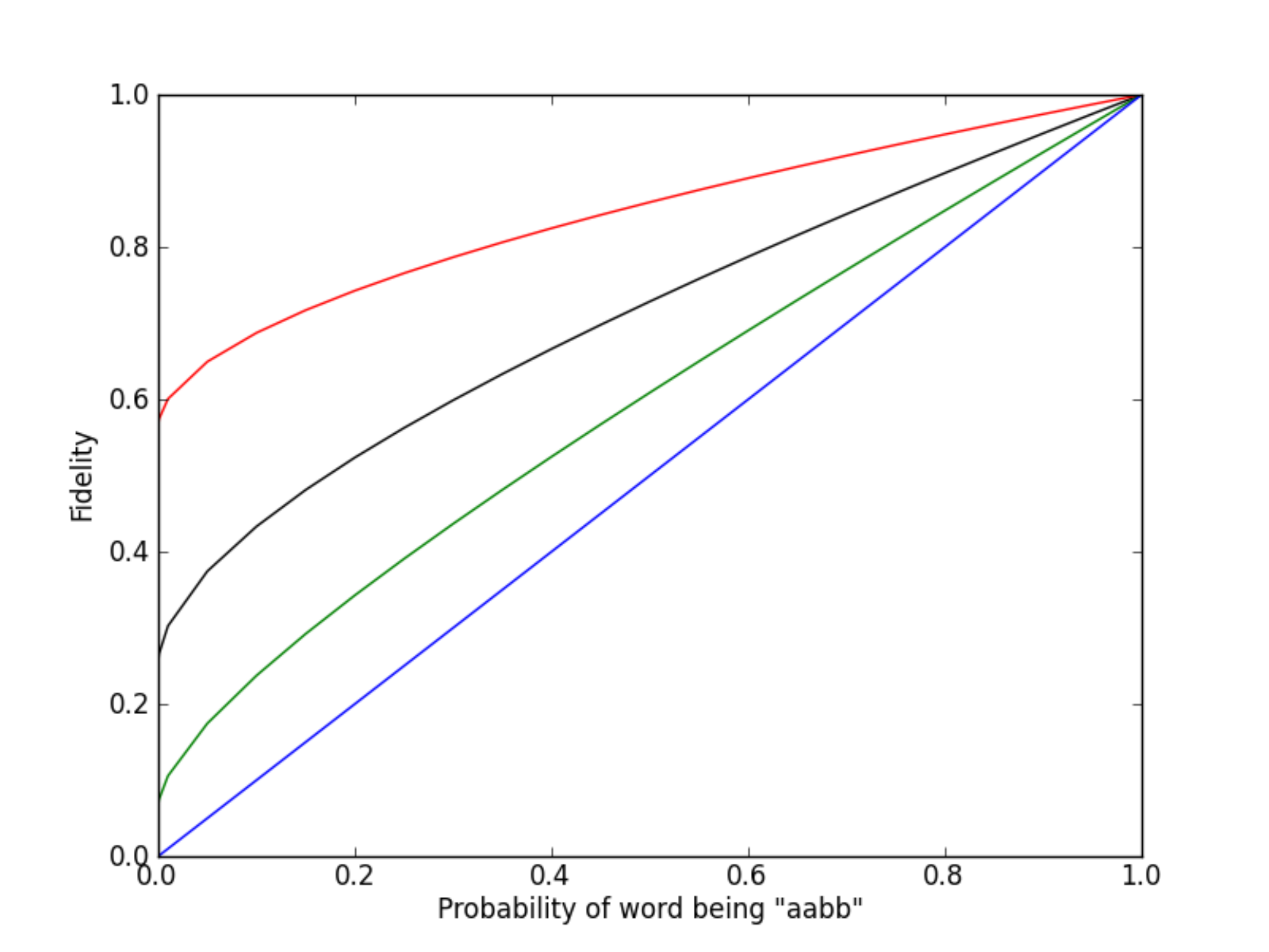}
\caption{Fidelity of final state to accepting state for quantum inputs in a superposition of $aabb$ and: the string with no matching characters, $bbaa$ (blue); the strings with one matching character, $abaa$, $baaa$, $bbab$ and $bbba$ (green); the strings with two matching characters $aaaa$, $abab$, $abba$, $baab$, $baba$ and $bbbb$ (black); the strings with three matching characters $aaab$, $aaba$, $abbb$ and $babb$ (red)} 
\label{fig:qinput}
\end{figure}

\textbf{Numerical study:}
as a preliminary investigation into using quantum inputs, we numerically tested the effects of using a quantum input consisting of a word accepted by the quantum walk superposed with a word not accepted.  We used the walks accepting $L_{eq}$ and tested the spatially distributed input as given in figure \ref{fig:ambmspa2} (a).  We used the fidelity between the state obtained and the accepting state to make our comparison.  The fidelity between states $| \psi \rangle $ and $| \phi \rangle $ is defined as:
\begin{equation}
F =|  \langle \phi | \psi \rangle |^2
\label{eq: fid}
\end{equation}
 We tested a range of superpositions from the word being entirely $aabb$ to the word being entirely another word, for every word of length four. The effect of the quantum input on the fidelity fell into four distinct cases, as shown in figure \ref{fig:qinput}. These cases correspond to the number of characters a word of length four can differ by, with the probability of being able to accurately determine which input was used diminishing as the number of symbols which match exactly between the two strings increases, as would be expected. 

Using a quantum input frames the question of language acceptance in terms of quantum state discrimination. If we know we have been given one of two states $| \psi \rangle $ and $| \phi \rangle $, one of which encodes a word in the language accepted by the walk while the other does not, then we can gain information about which state we are likely to have been given by whether or not it is accepted by the walk. Another approach to using discrete time quantum walks to perform quantum state discrimination, by measuring at specific positions, is presented in \cite{QWPOVM}.

%%%%%%%%%%%%%%%%%%%%%%%%%%%%%%%%%%%%%%%%%%%%%%%%%%%%%%%%%%%
\section{Summary}
\label{sec:sum}

We have presented a preliminary investigation into using discrete time quantum walks to recognise formal languages.  We developed two different types of graph structures, one using spatially distributed input, the other using sequential input.  We observed the expected trade off between space and time resources.  Spatially distributed input requires $O(n)$ vertices, but can recognise regular and context free languages in $O(1)$ steps of the quantum walk.  Sequential input requires only $O(1)$ vertices (besides those encoding the input) but uses $O(n)$ steps to recognise regular languages.  However, to recognise context free languages with sequential input requires $O(n)$ vertices, the extra vertices providing the necessary memory functionality.  

To gain insight into the behaviour of the quantum walks for input words not in the language, numerical testing was done for all possible inputs up to length 16.  The probability of acceptance was calculated and found to take a few specific values between 0 and 1, rather than varying over the entire range.  It would thus be interesting to classify which types of strings give rise to which values for the probability of acceptance. This could aid the process of finding walks that accept further languages. 

It would be interesting to extend the work to find general ways of specifying graphs accepting arbitrary regular expressions, or even arbitrary formal languages. Or one could restrict attention to specific, appropriately defined subclasses of regular expressions to relate the work to other aspects of logic. In order to compare the walks with other standard models of computation in terms of language acceptance, relations between the efficiency measures used here and other standard efficiency measures such as the minimal number of computational states, or tape squares traversed, must be found.

The two approaches discussed in this paper are unlikely to be the only ways to specify quantum walks such that they can be interpreted as accepting formal languages. It is not yet clear what the most fruitful approach will be. A better understanding of the relative merits of using spatially versus sequentially distributed inputs will be informative, and gaining insights into their limitations may suggest further ways of specifying walks to recognise languages. 

We have briefly described how the constructions presented here allow quantum inputs. This new form of input opens many further avenues for investigation. As well as finding out more about the quantum state discrimination performed by the walks presented here, it would be interesting to generalise the inputs themselves to superpositions of whole words rather than single symbols, and consider how they might be developed to form quantum languages.

%%%%%%%%%%%%%%%%%%%%%%%%%%%%%%%%%%%%%%%%%%%%%%%%%%%%%%%%%%%
\bibliographystyle{eptcs}

%\bibliography{bibfile}

\end{document}